\documentclass[%
 reprint,
 amsmath,amssymb,
 prd,
floatfix,
showkeys
]{revtex4-2}

\usepackage{graphicx}% Include figure files
\usepackage{dcolumn}% Align table columns on decimal point
\usepackage{bm}% bold math
\usepackage{hyperref}% add hypertext capabilities
\usepackage[mathlines]{lineno}% Enable numbering of text and display math
\usepackage{graphicx}
\usepackage{epstopdf}% To incorporate .eps illustrations using PDFLaTeX, etc.
\usepackage{subfigure}% Support for small, `sub' figures and tables
\usepackage{multirow}

\begin{document}

\title{Neutrino oscillation analysis of 217 live-days of Daya Bay and 500 live-days of RENO}% Force line breaks with \\

\author{Mario A.~Acero\textsuperscript{1}}
\email{marioacero@mail.uniatlantico.edu.co}
\author{Alexis A.~Aguilar-Arevalo\textsuperscript{2}}
\email{alexis@nucleares.unam.mx}
\author{Dairo J.~Polo-Toledo\textsuperscript{1}}
\email{djosepolo@mail.uniatlantico.edu.co}
\affiliation{
\textsuperscript{1} Universidad del Atl\'antico, Carrera 30 No.~8--49, Puerto Colombia, Atl\'antico, Colombia. \\
\textsuperscript{2} Instituto de Ciencias Nucleares, Universidad Nacional Aut\'onoma de M\'exico; CDMX 04510, M\'exico.
}

\date{\today}% It is always \today, today,
             %  but any date may be explicitly specified

\begin{abstract}
We present a neutrino oscillation analysis of two particular data sets from the Daya Bay and RENO reactor neutrino experiments aiming to study the increase in precision in the oscillation parameters $\sin^2{2\theta}_{13}$ and the effective mass splitting $\Delta m^2_{ee}$ gained by combining two relatively simple to reproduce analyses available in the literature. For Daya Bay the data from 217 days between December 2011 and July 2012 were used. For RENO we used the data from 500 live days between August 2011 and January 2012. We reproduce reasonably well the results of the individual analyses, both, rate-only and spectral, defining a suitable $\chi^2$ statistic for each case. Finally, we performed a combined spectral analysis and extract tighter constraints on the parameters, with an improved precision between 30-40\% with respect of the individual analyses considered.
\end{abstract}

\keywords{Neutrino oscillations; Reactors; Oscillation parameters}
\maketitle

\section{Introduction}
Since their discovery in 1956 \cite{reines:1956,Cowan:1992}, neutrinos have been under a heavy scrutiny by scientists trying to increase our knowledge about these abundant, exotic and enigmatic particles. 
Neutrinos are neutral, spin-{\scriptsize $\frac{1}{2}$}, weakly interacting particles which are found to exist in three different flavors: electron neutrinos ($\nu_e$), muon neutrinos $(\nu_{\mu})$ and tau neutrinos ($\nu_{\tau}$). According to the SM, neutrinos are massless particles, however, a variety of experiments carried out over the past 50 years have shown that they undergo a quantum mechanical interference phenomenon, known as neutrino oscillation \cite{Ghosh:2016jot}, through which the flavor of a neutrino changes while traveling from one point to another, implying that they must have  non-zero masses.
The discovery of neutrino oscillations was awarded the Nobel Prize in Physics in 2015.

Within the standard theory of neutrino oscillations, a neutrino of a given flavor can be expressed as a superposition of three definite-mass neutrinos $\nu_k$ ($k=1,2,3$) as
\begin{equation}\label{eq_nuState}
    \nu_{\alpha} = \sum_k U_{\alpha k} \;
    \nu_k,
\end{equation}
where $U_{\alpha k}$ are the elements of the Pontecorvo-Maki-Nakagawa-Sakata (PMNS) mixing matrix, which depend on the mixing angles $\theta_{kj}$ and a CP-violating phase $\delta_{CP}$ \cite{Giganti:2017fhf}. The PMNS matrix may also depend on two additional Majorana phases $\alpha_{1,2}$, which are not observable through neutrino oscillations.  The probability that a neutrino created with a given flavor $\nu_{\alpha}$ is detected as a different flavor $\nu_{\beta}$ after traveling a distance $L$ in vacuum is given by \cite{Giunti:2003qt,Giunti:2007ry}
\begin{widetext}
\begin{equation}\label{eq_OscProb}
    P_{\nu_{\alpha} \rightarrow \nu_{\beta}} = \delta_{\alpha\beta} 
     - 4\sum_{k>j}^3{\mathcal Re}\left[U^*_{\alpha k}U_{\beta k}U_{\alpha j}U^*_{\beta j}\right]\sin^2\left(\frac{\Delta m^2_{kj}L}{4E}\right)      + 2\sum_{k>j}^3{\mathcal Im}\left[U^*_{\alpha k}U_{\beta k}U_{\alpha j}U^*_{\beta j}\right]\sin\left(\frac{\Delta m^2_{kj}L}{2E}\right) ,
\end{equation}
\end{widetext}
where $E$ is the neutrino energy and $\Delta m_{kj}^2\equiv m_{k}^2 - m_{j}^2$ are the differences of the squared masses of the definite-mass states $k$ and $j$.
%---------------------------------
\begin{center}
\begin{figure*}
\resizebox*{17.0cm}{!}{\includegraphics{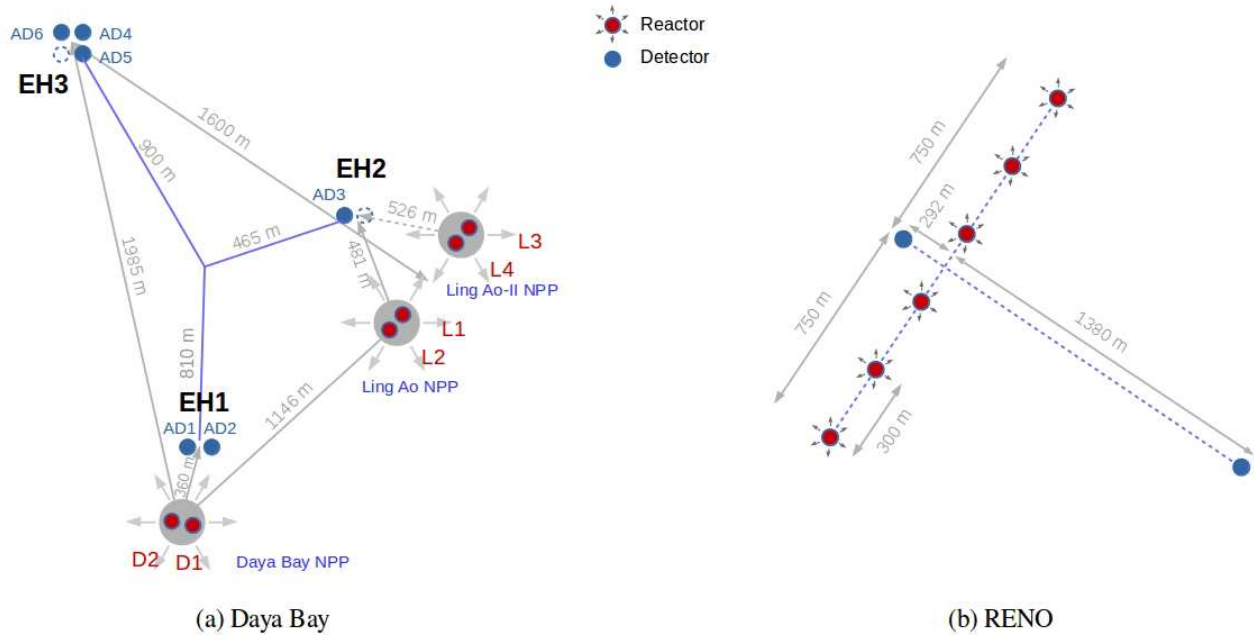}}
\caption{Arrangement of the nuclear reactors, shown as red circles and antineutrino detectors (AD), shown as blue circles, for the Daya Bay (6 AD configuration) and RENO reactor neutrino experiments. Adapted from \cite{Mezzetto:2010}.}
\label{fig_sites}
\end{figure*}
\end{center}
%---------------------------------

In the period from the late 1990's to early 2010's, definitive experimental confirmation of neutrino oscillations was gathered from atmospheric \cite{Fukuda:1998mi,Kajita:2010zz}, solar (e.g. SNO \cite{Ahmad:2002jz,Aharmim:2011vm}), long baseline accelerator (K2K \cite{Ahn:2006zza}, MINOS \cite{Michael:2006rx}, T2K \cite{Abe:2013hdq}), and very-long baseline reactor (KamLAND \cite{Araki:2004mb}) neutrino experiments. In 2012 the long baseline reactor neutrino experiments Daya Bay \cite{An:2012eh}, RENO \cite{Ahn:2012nd}, and Double Chooz \cite{Abe:2013sxa}, reported the first measurement of the mixing angle $\theta_{13}$, by observing the disappearance of reactor antineutrinos ($\bar{\nu}_e$) over distances of the order of 1 km, finding a non-zero value, and opening the door to studying CP violation in the neutrino sector. In later years, neutrino oscillations research has entered into a precision era, where the oscillation parameters can be determined with percent level precision from analyses that combine the results of many different experiments \cite{deSalas:2017kay,Esteban:2018azc}. The current focus of the field is primarily oriented to the determination of the CP--violating phase $\delta_{CP}$, the neutrino mass ordering, and the octant of the angle $\theta_{23}$ (see for instance \cite{Acero:2019ksn} and \cite{Abe:2018wpn} for recent experimental results, and \cite{HyperK:2018,Abi:2018dnh} for current progress on new experimental efforts).

In this work we perform a combined analysis of the data from two specific data-taking periods of the Daya Bay and RENO experiments. For Daya Bay we consider the 217 days of data, taken between December of 2011 and July of 2012 in the configuration with only 6 antineutrino detectors \cite{An:2013zwz}. In the case of RENO, we consider the data from 500 live days, taken between August 2011 and January 2013 \cite{Seo:2016uom} with both, the near and far detectors. These data sets have been chosen for the convenience and relative simplicity in reproducing their results from publicly available resources. Although more recent data are available, we have not considered them here. The aim of this work is to study the level of precision that can be attained by combining such older data sets, as well as to test our reproduction of the Daya Bay result with a full covariance matrix approach. In the following sections we provide information about the experiments as well as a description of our analysis and results.
%******************************************************************************
%******************************************************************************
\section{Antineutrinos from nuclear reactors}\label{reactorNus}
Typical commercial pressurized water reactors (PWR) are copious sources of $\bar{\nu}_e$s, originating primarily in the beta decay of unstable fission products of the fissile isotopes $^{235}$U, $^{238}$U, $^{239}$Pu, and $^{241}$Pu, as well as from neutron capture in $^{238}$U. On average, each fission releases roughly $200$~MeV of energy and produces $6$ antineutrinos with energies below 10 MeV. Since the typical decay chain of the fission products has three consecutive beta decays, about $2\times 10^{20}$ $\bar{\nu}_e$ per second per Giga--Watt of thermal power (GW$_{\rm{th}}$) \cite{Giunti:2007ry} are isotropically emitted from the reactor core.
The neutron capture contribution occurs at a smaller rate (0.6 per fission) and produces ${\bar\nu}_e$s with energies below 1.3~MeV \cite{Wong:2006}.

Reactor antineutrinos with energies $>1.8$~MeV can be detected via the inverse beta decay (IBD) process 
\begin{equation}\label{eq_IBD}
    \bar{\nu}_e + p \rightarrow e^+ + n,
\end{equation}
by recording the delayed coincidence of the positron and neutron capture signals in, for example, a scintillating detector doped with a high neutron capture cross section material, like Gadolinium (Gd). The convolution of the reactor antineutrino flux and the IBD cross section gives an energy spectrum of the detected ${\bar\nu}_e$s with a peak around 3~MeV, and a cutoff at the 1.8~MeV threshold of the reaction.

After the initial observation by RENO of a feature in the the ${\bar\nu}_e$ spectrum that has come to be known as the  ``5~MeV-bump'', and its subsequent confirmation by Daya Bay, Double-Chooz, and other experiments, significant interest has arisen to try to explain it within the boundaries of nuclear physics, as well as through non-standard particle physics (see \cite{berryman:2019,huber:2016,huber:2017,buck:2017,dwyer:2016} and references therein). The oscillation analyses developed by RENO and Daya Bay, which we reproduce here, assume that the bump is unrelated to the physics of neutrino oscillations, and are mostly unaffected by this feature.

\subsection{Brief description of the experiments}\label{DB}
The Daya Bay experiment is located nearly 55 km northwest of Hong Kong, it uses the antineutrinos emitted by six functionally identical PWRs of 2.9~GW$_{\rm th}$ each \cite{An:2012eh}, two of them located in the Daya Bay Nuclear Power Plant (NPP), and four in the neighboring Ling Ao and Ling Ao--II NPPs. In Figure \ref{fig_sites}(a) the red circles represent the nuclear reactors, arranged in pairs (2 in Ling Ao, 2 in Ling Ao--II, and 2 in Daya Bay), and the blue circles represent the antineutrino detectors (AD).
In the data taking period used in this work, the detectors were distributed in three experimental halls (EH1, EH2, EH3) as depicted in the figure. The three EHs are, respectively, under 250, 265, and 860 m.w.e. of overburden, and are interconnected through internal tunnels, in order to shield the detectors from cosmic rays and other sources of radiation. The full 8 AD configuration was completed in 2012. Further details can be found in \cite{An:2015qga}.

The Reactor Experiment for Neutrino Oscillation (RENO) is located in the Hanbit (formerly Yonggwang) NPP in the southwest coast of South Korea, 250 km south of Seoul \cite{Ahn:2010vy}, and uses the $\bar{\nu}_e$ from six PWRs arranged in a line along the coast. The reactors produce a total of 16.4 GW$_{\rm{th}}$. RENO uses two detectors (Near Detector --ND-- and Far Detector --FD--) to observe the produced $\bar{\nu}_e$s, as displayed in Figure \ref{fig_sites}(b), where the red circles represent the six reactors, and the blue circles represents the two detectors. The ND (FD) is under 120 (450) m.w.e.~of overburden \cite{Kim:2016}. The average distance from the reactors to the ND (FD) is 292~m (1380~m) \cite{Seo:2016uom}.

%---------------------------------
\begin{center}
\begin{figure}
\centering
\resizebox*{7.0cm}{!}{\includegraphics{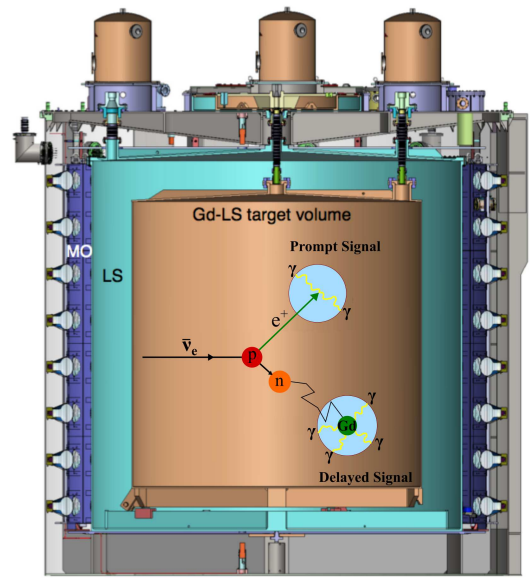}}\\
\vspace{0.3cm}
\centering
\resizebox*{7.0cm}{!}{\includegraphics{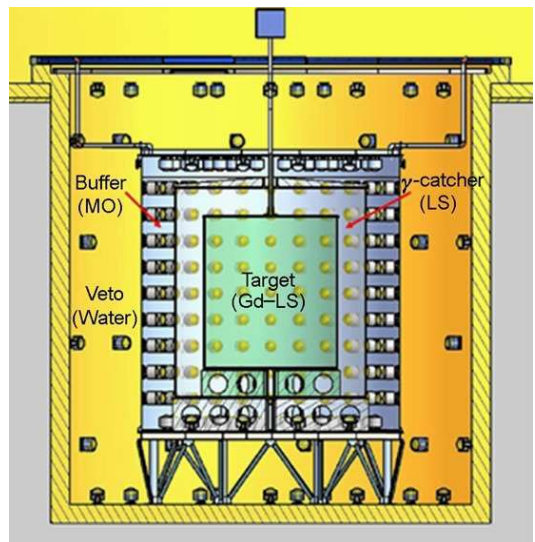}}
\caption{Schematic view of the Daya Bay ({\it top}, adapted from \cite{Heeger:2011}) and RENO ({\it bottom} \cite{Kim:2016}) antineutrino detectors, with the three concentric vessels (containing Gd-LS, LS and MO) clearly identified. The interaction of a $\bar{\nu}_e$ with a proton in the target via IBD, is shown inside the Daya Bay inner-most cylinder.} \label{fig_detector}
\end{figure}
\end{center}
%---------------------------------

Daya Bay and RENO observe $\bar{\nu}_e$’s through the IBD reaction, Eq.~(\ref{eq_IBD}).
Both experiments use a similar detector design with three concentric cylinders containing different liquids (see Figure \ref{fig_detector}). The inner-most cylinder is filled with a Gd-doped liquid scintillator (LS) and acts as the main target volume; the intermediate one, designed to efficiently detect gamma rays (gamma catcher), is filled with pure LS, and the outer one, whose inner walls are lined with photomultiplier tubes (PMTs), is filled with mineral oil, which acts as a buffer. The detectors are immersed in water pools whose walls are instrumented with PMTs and work as vetoes. Details of the detector design of each experiment can be found in \cite{An:2015qga} for DB and \cite{Park:2012} for RENO.

In the target volume, a large amount of freely moving protons $(p)$ may interact with the antineutrinos coming from the reactors, producing positrons $(e^+)$, which then annihilate with surrounding electrons generating two gamma rays (prompt signal). In addition, in the IBD process, a neutron $(n)$ is also created; this thermalizes and is captured by a Gd nucleus, emitting more gamma rays (delayed signal). The time difference between these two signals is a few $\mu s$. A representation of the particle identification signal is shown inside the Daya Bay detector in the top panel of Figure \ref{fig_detector}.
Once the detected signals are collected, specialized selection criteria are applied by the experiments to estimate the observed number of $\bar{\nu}_e$ events and background rates in each detector (for detailed information about this process see, for instance, \cite{An:2012eh,An:2013uza,Ahn:2012nd}).

\subsection{Input to our studies}
\label{sec:inputs}
\begin{center}
\begin{figure*}
    \begin{minipage}{0.99\columnwidth}
        \centering %\hspace*{-4.0cm}
        \includegraphics[scale=0.43]{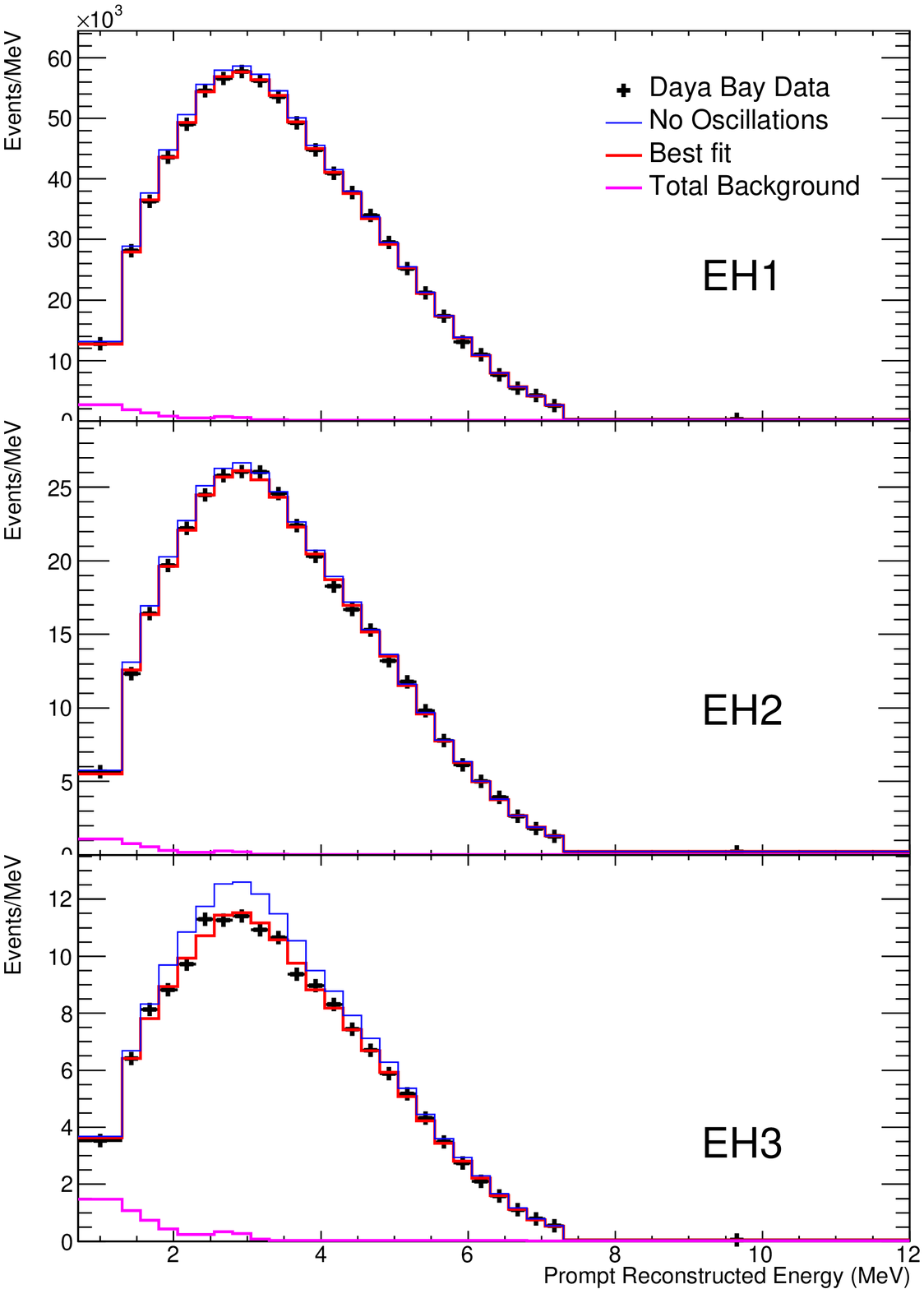}
    \end{minipage} \hspace{-0.0cm}
    \begin{minipage}{0.99\columnwidth}
        \centering
        \includegraphics[scale=0.43]{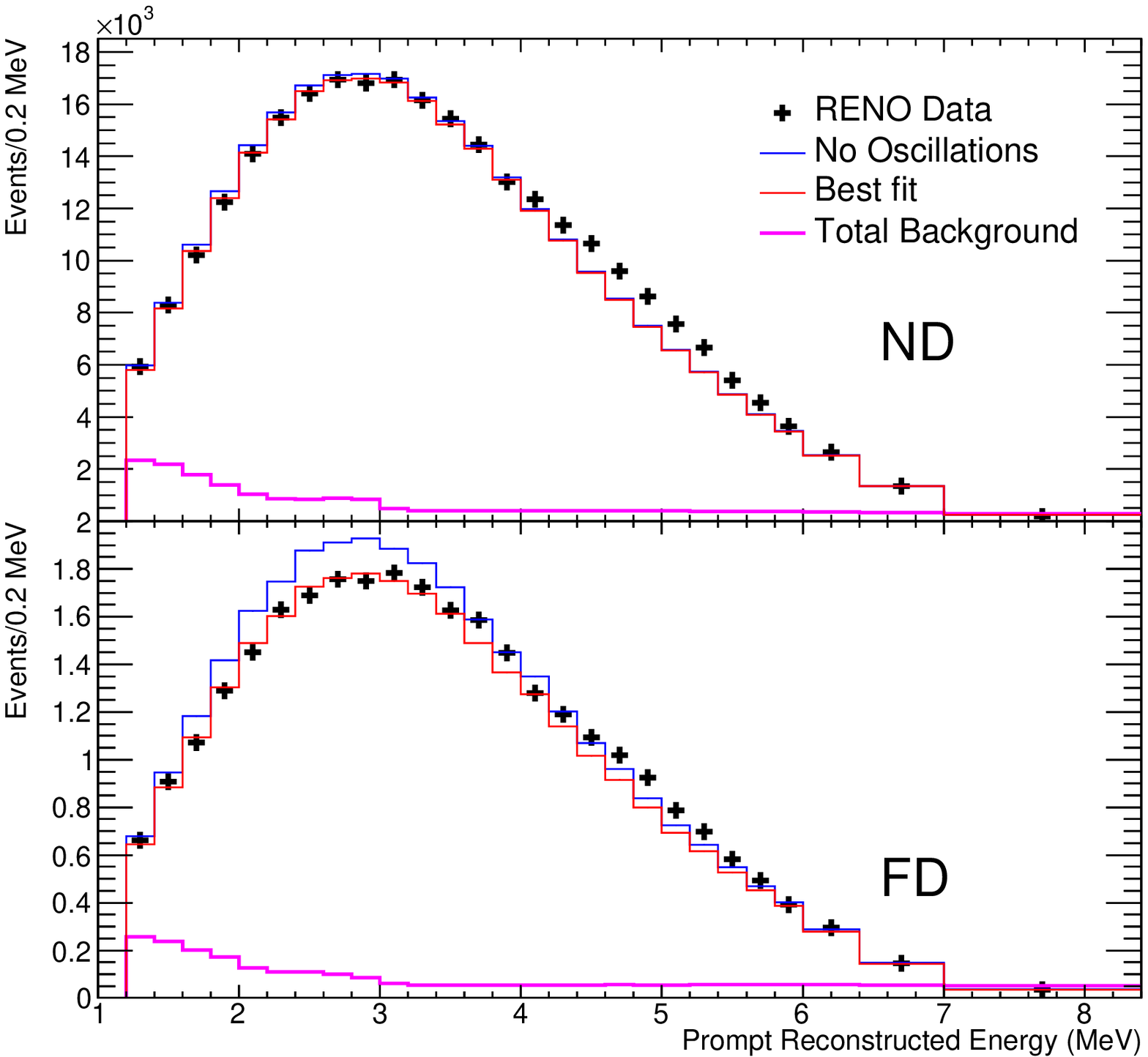}
        \\ \vspace{-0.0cm}
        \includegraphics[scale=0.43]{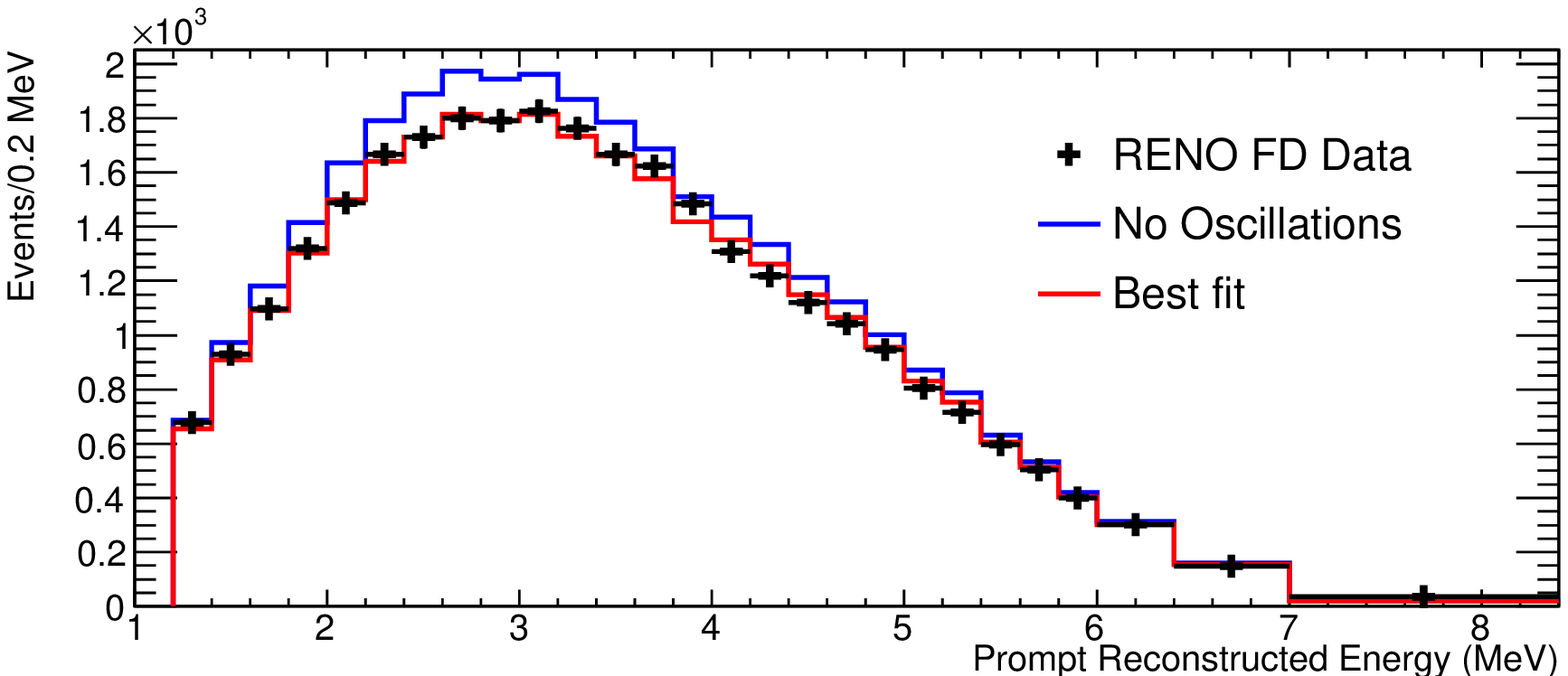}
    \end{minipage}
\caption{Background-subtracted reconstructed positron prompt energy distributions for Daya Bay (left) and RENO (right). Black crosses are data, blue (red) line histograms show the MC expected no-oscillation (best fit) spectra. The background histograms are shown in magenta. See text for details.
For Daya Bay the distributions are absolutely normalized and account for the 5~MeV bump effect. For RENO the predicted ND and FD distributions are normalized such that the best fit and data histograms have equal areas outside the 5~MeV bump region ($3.8~{\rm MeV}<E_p<6.4~{\rm MeV}$). The bottom-right plot (RENO) compares the observed spectrum in the FD with no-oscillation and the best fit predictions obtained from the measurement at the ND, which account for the effect of the 5~MeV bump. 
} \label{fig_spectra} 
\end{figure*}
\end{center}

The prompt reconstructed energy, $E_p$, distributions used in the analyses are shown in Figures \ref{fig_spectra}. For Daya Bay, we digitized the data and no-oscillation distributions from figure 2 in Ref.~\cite{An:2013zwz}, and assume that all detectors in the same EH have the same distribution. Note that the predicted Daya Bay distributions already account for the 5~MeV bump. For RENO, we digitized the data and best-fit distributions from figure 26 in Ref.~\cite{Seo:2016uom}. The no-oscillation distributions in the near and far RENO detectors were constructed by removing from the best-fit spectrum, bin by bin, the effect of the oscillations with the help of a sample of simulated neutrino events (see section \ref{sec:simulation}). Note that in the RENO case, the predicted distributions do not include the 5~MeV bump; however, their spectral analysis, based on a far-to-near ratio, described later, will prove to be insensitive to this effect.
%---------------------------------
%---------------------------------

In order to normalize the event rates and energy distributions, we collected the information from tables found in Refs. \cite{An:2013zwz} and \cite{Seo:2016uom}, which we have summarized here in Table \ref{tab_DBdata} and Table \ref{tab_RENOdata}, for Daya Bay and RENO, respectively. In these tables we have added the estimates of the total IBD rates without oscillations used in our simulation for each experiment. In our RENO simulation, we set the total predicted IBD rate at the best fit to the observed value, and used the approximation of a common detection efficiency for the near and far detectors. In addition, for the Daya Bay spectral analysis, we digitized the full systematic error correlation matrix from Ref.~\cite{Wong:2016}, the total systematic errors from figure 2 in Ref.~\cite{An:2015nua}, and used this information to construct the full covariance matrix, as will be described in section \ref{analysis} below.

\subsection{Simulation of neutrino events} \label{sec:simulation}
\begin{center}
\begin{figure}[t!]
\centering
\resizebox*{8.5cm}{!}{\includegraphics{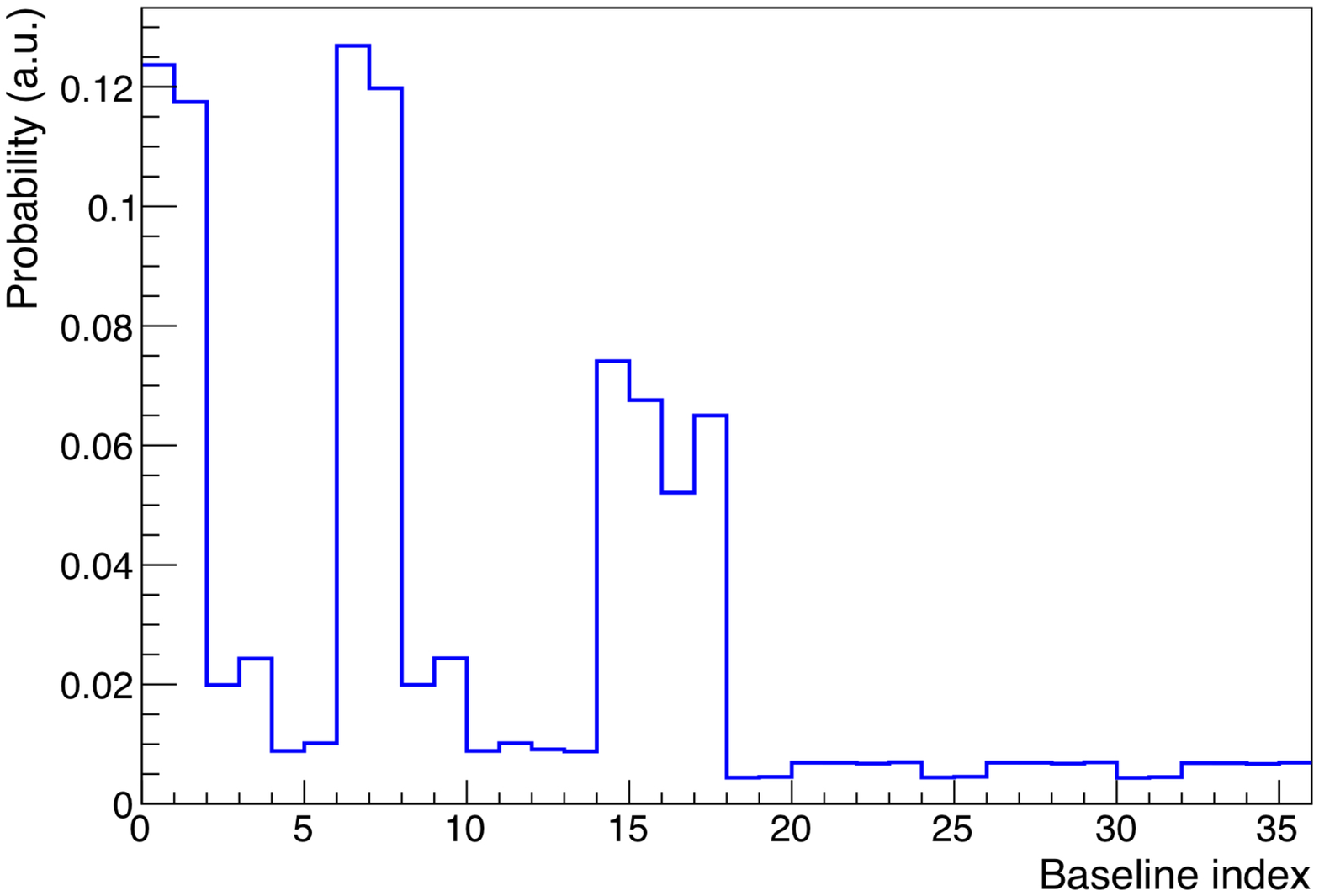}}\vspace{5pt}
\resizebox*{8.5cm}{!}{\includegraphics{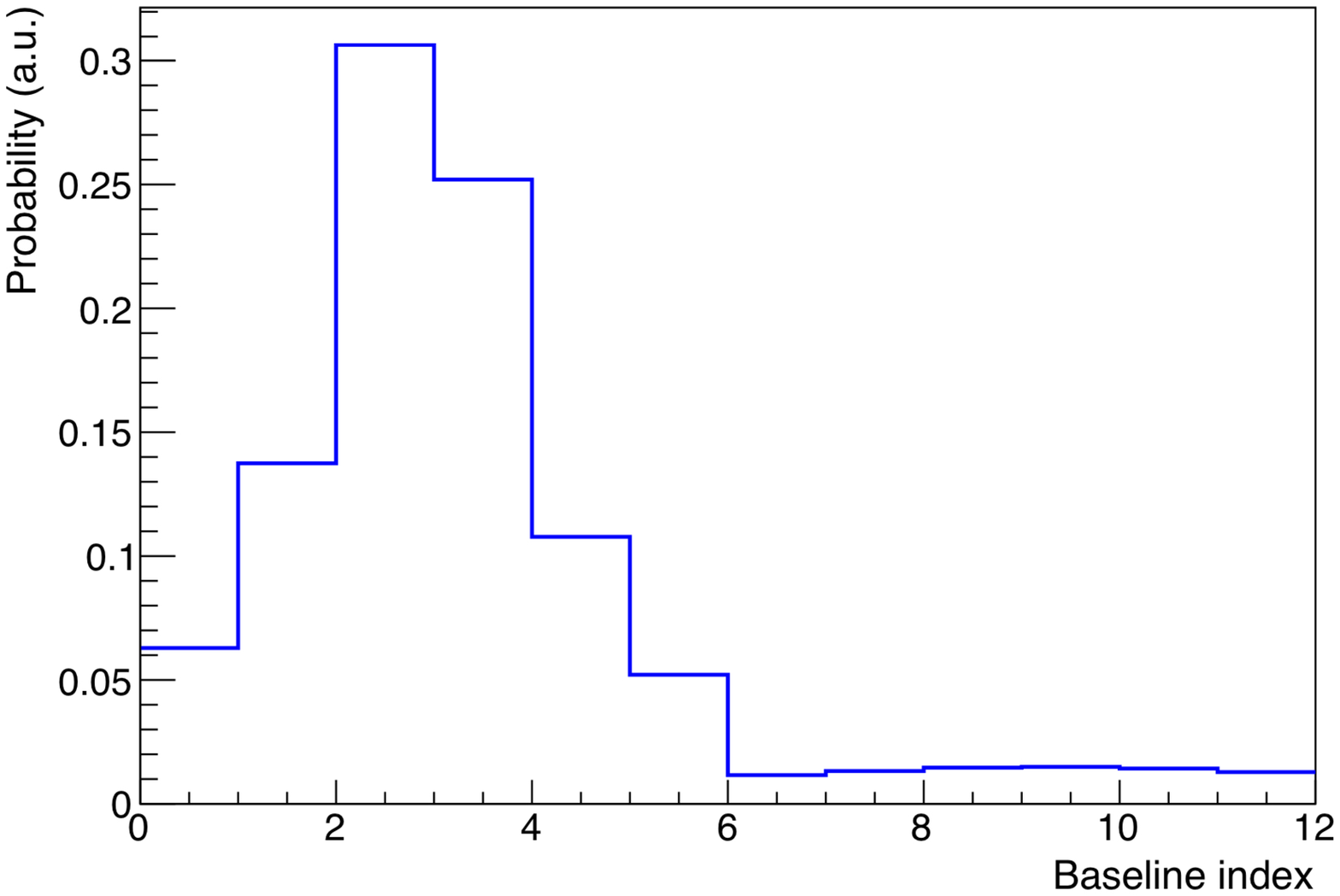}}
\caption{Baseline probability distributions for Daya Bay (top) and RENO (bottom), obtained from the simulation of $10^7$ neutrino events. Histograms are normalized to unit area.} \label{fig_ldist}
\end{figure}
\end{center}

We simulate neutrino events traveling the different baselines available between the various reactors and detectors in each experiment by constructing the probability that a neutrino leaves a particular reactor $r$ and arrives at a specific detector $d$. For RENO (6 reactors and 2 detectors), there are 12 different baselines, while for Daya Bay (6 detectors and 6 reactors) there are 36 different baselines, in the 6 AD configuration considered here. This probability is calculated as follows:
\begin{equation}
    w_r^d = N \frac{P_r^{th}M_d}{4\pi L_{rd}^2}, 
\end{equation}
where $P_r^{th}$ is the thermal power of reactor $r$, $M_d$ is the mass of the fiducial volume of detector $d$, and $L_{rd}$ is the baseline distance between reactor $r$ and detector $d$; $N$ is a normalization constant making the sum $\sum_{rd}w_{r}^d = 1$.
The baseline lengths and detector fiducial volume masses were extracted from Ref.~\cite{An:2015qga} for the Daya Bay analysis, and from Ref.~\cite{Ahn:2010vy} for the RENO analysis. Figure \ref{fig_ldist} shows the probability distributions for a neutrino to travel along each available baseline. Note that each baseline index (1-12 for RENO, and 1-36 for Daya Bay) uniquely identifies a reactor-detector pair. The histograms give the probability that a neutrino in the experiment was produced in a particular reactor and observed in a particular detector. As expected, near detectors have larger probabilities than far detectors, since the closer the detector is to a reactor, the more antineutrinos are detected.
%---------------------------------
\begin{center}
\begin{table*}
\caption{Daya Bay IBD candidates and expected IBD rates and backgrounds per day in the 217-day sample used in this work. The DAQ live time and the muon veto and multiplicity-cut efficiencies ($\epsilon_\mu\cdot\epsilon_m$) are also reported for each AD. The last line shows the expected IBD rate without oscillations. Adapted from \cite{Abe:2013hdq}.}
\begin{ruledtabular}
\begin{tabular}{lcccccc} %\toprule
 Experimental Hall & \multicolumn{2}{c}{EH1} & \multicolumn{2}{c}{EH2} & \multicolumn{2}{c}{EH3} \\ \colrule
 Detector & AD1 & AD2 & AD3 & AD4 & AD5 & AD6 \\ \colrule
 IBD Candidates/day & $530.21\pm1.67$ & $536.75\pm1.68$ & $489.93\pm1.61$ & $73.58\pm0.62$ & $73.21\pm0.62$ & $72.35\pm0.62$ \\
 DAQ live time (days) & \multicolumn{2}{c}{191.001} & \multicolumn{2}{c}{189.645} & \multicolumn{2}{c}{189.779} \\
 $\epsilon_\mu\cdot\epsilon_m$ & 0.7957 & 0.7927 & 0.8282 & 0.9577 & 0.95689 & 0.9566 \\ 
 Total background/day & $13.20\pm0.98$ & $13.01\pm0.98$ & $9.57\pm0.71$ & $3.52\pm0.14$ & $3.48\pm0.14$ & $3.43\pm0.14$ \\
 IBD Rates/day (best fit) & $653.30\pm2.31$ & $664.15\pm2.33$ & $581.97\pm2.07$ & $73.31\pm0.66$ & $73.03\pm0.66$ & $72.20\pm0.66$ \\ 
 No-osc.~IBD Rates/day & $664.72$ & $675.56$ & $593.57$ & $79.05$ & $78.75$ & $77.85$
\end{tabular}
\end{ruledtabular}
\label{tab_DBdata}
\end{table*}
\end{center}
%---------------------------------
%---------------------------------
\begin{center}
\begin{table}[t!]
\caption{RENO IBD rates and backgrounds per day in the 500 live days used in this work. DAQ live times are also reported for the Near and Far detectors. A common detection efficiency was used in our simulation. Adapted from \cite{RENO:2015ksa}.}
\begin{ruledtabular}
\begin{tabular}{lcc} %\toprule
Detector & Near & Far \\ \colrule
 IBD Rate (background sub.) & $616.67\pm1.44$ & $61.24\pm0.42$ \\
 DAQ live time (days) & 458.49 & 489.93 \\  
 Detection efficiency &  0.7644 & 0.7644 \\
 Total background & $17.54\pm0.83$ & $3.14\pm0.23$ \\
 No-osc.~IBD Rate & $623.26$ & $65.57$ %\\ \colrule
\end{tabular}
\end{ruledtabular}
\label{tab_RENOdata}
\end{table}
\end{center}
%---------------------------------

The neutrino energy for each simulated event is then assigned using the well-known relation \cite{Abe:2013hdq}
\begin{equation}\label{eq_nuEnergy}
    E_{\nu} = E_p + \overline{E}_n + 0.78 \, \rm{MeV},
\end{equation}
where $\overline{E}_n$ is the average energy taken by the neutron ($\sim 10$ keV). A small multiplicative correction factor $f=1.03$ was applied to $E_p$ ($E_p\rightarrow E_p\times f$) in our RENO simulation to better reproduce the results for this experiment. Since we do not have access to the no-oscillated RENO spectra, this factor, which is degenerate with $\Delta m_{ee}^2$, serves the purpose to make our overly simplistic simulation resemble more closely that developed by the collaboration. While this neutrino energy is rather a reconstructed quantity and not the actual true energy of the event, the energy resolution effect will be neglected, as it is reported to be smaller than the bin size.

As mentioned in section \ref{sec:inputs}, Refs.~\cite{An:2013zwz} and \cite{Seo:2016uom} only report the number of either expected or observed IBD events in each detector, with the effect of oscillations at the best-fit. However, knowledge of the number of IBD candidate events expected without oscillations is required. 
We estimated the number of events without oscillations dividing the number of oscillated events in a given bin, or full spectrum, by the average best-fit oscillation probability of all the events in said bin or spectrum. Average oscillation probabilities were calculated by applying the oscillation probability in Eq.~(\ref{eq_survProb_reduced}) to a sample of $10^7$ simulated neutrino IBD events whose \emph{true} energy $E_{\nu}$, prompt positron energy $E_p$, and baseline $L$, are assigned as follows: first, a baseline $L$ is randomly sampled from the distribution of baselines in Figure \ref{fig_ldist}, this uniquely determines the reactor-detector pair for the event. A random Gaussian fluctuation ($\sigma=1$~m) is added to approximately incorporate the reactor and detector sizes. The prompt positron energy %$E_p$ 
is then sampled from the $E_p$ distribution corresponding to the chosen detector (and corrected by multiplying it by $f$). This $E_p$ value is then used to calculate the {\it true} neutrino energy $E_\nu$ using Eq.~(\ref{eq_nuEnergy}).
%---------------------------------
%---------------------------------
%******************************************************************************

%******************************************************************************
\section{Oscillation analysis}\label{analysis}
Neutrino oscillations in long baseline reactor neutrino experiments manifest themselves as the disappearance of electron antineutrinos with energies between 2 and 6 MeV over distances of the order of 1~km. In this case the survival probability $P_{\bar{\nu}_e \rightarrow \bar{\nu}_e}$ in Eq.~(\ref{eq_OscProb}) is well approximated by
\begin{align}\label{eq_survProb}
    P_{\bar{\nu}_e \rightarrow \bar{\nu}_e} &= 1 - \cos^4{\theta_{13}} \sin^2{2\theta_{12}} \sin^2{\Delta_{21}} \\
    & - \sin^2{2\theta_{13}} \left(\cos^2{\theta_{12}} \sin^2{\Delta_{31}} + \sin^2{\theta_{12}} \sin^2{\Delta_{32}} \right), \nonumber
\end{align}
with $\Delta_{ij} = 1.267\,\Delta m_{ij}^2\,L/E_{\nu}$, and 
$\Delta m_{ij}^2$ is given in eV$^2$, $L$ (in m) is the distance between the reactor and the detector, and $E_{\nu}$ (in MeV) is the neutrino energy. Given that $\Delta m_{21}^2 \ll \left|\Delta m_{31}^2\right| \approx \left|\Delta m_{32}^2\right|$, the $\bar{\nu}_e$ oscillation is mainly driven by $\Delta_{31}$, and Eq.~(\ref{eq_survProb}) naturally leads to the definition of an effective squared-mass difference $\Delta m^2_{ee}$ such that $\sin^2{\Delta_{ee}} = \cos^2{\theta_{12}}\sin^2{\Delta_{31}} + \sin^2{\theta_{12}}\sin^2{\Delta_{32}}$ \cite{An:2013zwz}, so that the survival probability in  Eq.~(\ref{eq_survProb}) can be reduced to
\begin{align}\label{eq_survProb_reduced}
    P_{\bar{\nu}_e \rightarrow \bar{\nu}_e} = 1 
    & - \cos^4{\theta_{13}} \sin^2{2\theta_{12}} \sin^2{\Delta_{21}} \nonumber \\
    & - \sin^2{2\theta_{13}} \sin^2{\Delta_{ee}}.
\end{align}
For the purpose of reproducing the Daya Bay results, in what follows we have used  $\sin^22\theta_{12}=0.857$, $\Delta m^2_{21}=7.50\times10^{-5}~{\rm eV}^2$, for the spectral analysis, and also $|\Delta m^2_{32}| = 2.32\times10^{-3}\,{\rm eV}^2$ for the rate only analysis, as in Ref.~\cite{An:2013zwz}. For the reproduction of the RENO results we used $\sin^22\theta_{12}=0.846$, and $\Delta m^2_{21}=7.53\times10^{-5}~{\rm eV}^2$, for the spectral analysis, and in addition $|\Delta m^2_{32}| = 2.49\times10^{-3}\,{\rm eV}^2$ for the rate only analysis, as in Ref.~\cite{Seo:2016uom}.

It has been pointed out that the definition of $\Delta m^2_{ee}$ used here, besides being $L/E$ dependent, is discontinuous at 0.5 km/MeV \cite{Parke:2016}, and better definition can be considered, such as the weighed average of $\Delta m^2_{31}$ and $\Delta m^2_{32}$. In the interest of attempting to reproduce the original results by Daya Bay and RENO, we will keep the definition in Ref.~\cite{An:2013zwz}, as this has not critical impact for our analysis.

We now present the procedure to estimate the oscillation parameters which best fit the data, preforming two types of approaches: a rate-only analysis and a spectral analysis.

\subsection{Rate-only analysis}
Using only the information of the total event rates reported in Table \ref{tab_DBdata} and Table \ref{tab_RENOdata}, and the measured values of the oscillation parameters $\sin^2{2\theta_{12}}$, $\Delta m_{21}^2$ and $\Delta m_{32}^2$, we extract the value of $\sin^2 2\theta_{13}$. Since the shape of the spectrum is not considered, no information about $\Delta m_{ee}^2$ is obtained from this analysis.

For Daya Bay, we follow Ref.~\cite{An:2012eh} and define our $\chi^2$ statistic as:
\begin{align}\label{eq_chi2DB_rate}
    \chi^2 = &\sum_{d=1}^{6} \frac{\left[M_d - T_d\left( 1 + \varepsilon + \sum_{r}\omega_r^d\alpha_r + \varepsilon_d\right) + \eta_d\right]^2}{M_d+B_d} \nonumber \\
    & + \sum_{r}\left(\frac{\alpha_r^2}{\sigma_r^2}\right) + \sum_{d=1}^{6}\left(\frac{\varepsilon_d^2}{\sigma_d^2} + \frac{\eta_d^2}{\sigma_B^2}\right).
\end{align}
Here, $M_d$ is the number of IBD observed events in the $d$-th detector after background subtraction, $B_d$ is the background rate, and $T_d$ is the predicted number of events considering neutrino oscillations through Eq.~(\ref{eq_survProb_reduced}), in the total DAQ live time (Table \ref{tab_DBdata}). The quantity $\omega_r^d$ is the fraction of events produced in reactor $r$ which contribute to detector $d$, considering the travel distance and the neutrino flux, which corresponds precisely with the baseline probability of Figure \ref{fig_ldist}. The $\chi^2$ in (\ref{eq_chi2DB_rate}) is penalized by the inclusion of 18 pull terms $\alpha_r$, $\varepsilon_d$, and $\eta_d$ (with $r,d=1,\dots,6$), characterizing the systematic errors affecting the measurement.  As reported by the collaboration in Ref.~\cite{An:2012eh}, $\sigma_{r}$ (0.8\%) and $\sigma_d$ (0.2\%) are the uncorrelated reactor and detector uncertainties, respectively, and $\sigma_B$ is the corresponding background uncertainty. The additional parameter $\varepsilon$ is included as a normalization factor which accounts for possible differences between the observation and the prediction, and it is included as a free parameter.

The interval $0<\sin^22\theta_{13}<0.2$ is split in 200 uniform steps. For each point a full minimization over the 18 pull terms and the parameter $\varepsilon$ is performed to obtain the value of the marginalized $\chi^2$ statistic. Minimization of the marginalized $\chi^2$ gives the best-fit value  $\sin^2{2\theta}_{13} = 0.090_{-0.009}^{+0.010}$ at 1$\sigma$ C.L.
%---------------------------------
\begin{center}
\begin{figure*}[t]
\centering
{
\resizebox*{15cm}{!}{\includegraphics{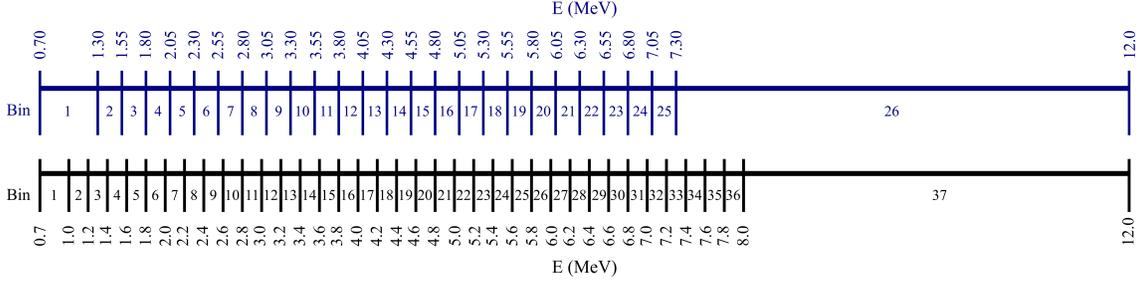}}}
\caption{Energy range and bin boundaries for the Daya Bay analysis in \cite{Wong:2016} (black lines and numbers), and for the analysis presented in this work (blue lines and numbers).
} \label{fig_rebin}
\end{figure*}
\end{center}
%---------------------------------

A similar rate-only analysis was performed to the RENO data, considering the three-neutrino oscillation model described by Eq.~(\ref{eq_survProb_reduced}). In this case, we follow Ref.~\cite{Ahn:2012nd} and define the $\chi^2$ statistic as
\begin{align}\label{eq_chi2RENO_rate}
    \chi^2 =& 
    \sum_{d=N,F}\frac{\left[ N_{obs}^d + b_d - \left( 1 + a + \xi_d \right) \sum_{r}\left(1 + f_r\right)N_{\exp}^{d,r}\right]^2}{N_{obs}} \nonumber \\
    &+ \sum_{d=N,F}\left(\frac{\xi_d^2}{(\sigma_d^{\xi})^2} + \frac{b_d^2}{(\sigma_d^{b})^2}\right) + \sum_{r=1}^{6}\left(\frac{f_r}{\sigma_r}\right)^2,
\end{align}
where $N_{obs}^d$ is the number of observed events after background subtraction for the near ($N$) and far ($F$) detectors in the total DAQ live time (Table \ref{tab_RENOdata}); $N_{\exp}^{d,r}$ is the number of expected events in detector $d$ coming from reactor $r$, including the detection efficiency and the effect of oscillations. Here $a$ is a freely varying normalization factor, $\sigma_b^d$, from Table \ref{tab_RENOdata} are the background uncertainties associated to the pull term parameters $b_d$, and $\sigma_{r}$ (0.9\%) and $\sigma_{d}^\xi$ (0.2\%) are the uncorrelated reactor and detector uncertainties, associated to the pull term parameters $f_r$ and $\xi_d$, respectively. Using a similar minimization procedure to the one used for the Daya Bay analysis, marginalizing over $a$ and the pull term parameters, we obtain $\sin^2{2\theta}_{13} = 0.088_{-0.013}^{+0.010}$ at 1$\sigma$ as the value which best fit to the RENO data.

\subsection{Spectral analysis}
Here, besides the normalization information, the shape of energy distribution of the observed IBD events will be used to extract the value of $\Delta m_{ee}^2$ along with $\sin^2{2\theta}_{13}$. 
The best fit is found by minimizing a suitable $\chi^2$ statistic defined over a uniformly spaced 100$\times$100 grid in the $\sin^22\theta_{13}$ {\it vs} $\Delta m^2_{ee}$ space. 

\subsubsection*{Daya Bay}
For Daya Bay we follow Ref.~\cite{An:2016srz} and define
\begin{equation}\label{eq_chi2DB_spec}
    \chi^2 = \sum_{i,j} \left(N_i^{obs} - N_i^{exp}\right)^T V_{ij}^{-1} 
    \left(N_j^{obs} - N_j^{exp}\right),
\end{equation}
where the indices $i$ and $j$ run over 156 bins corresponding to the concatenation of the prompt energy distributions of the 6 ADs, with 26 bins each. 
$N_i^{obs}$ ($N_i^{exp}$) is the number of observed (expected) events in the $i$-th energy bin, and $V_{ij}$ are the elements of the total covariance matrix expressed in the same binning. $N_i^{exp}$ depends on the oscillation parameters $\sin^2{2\theta}_{13}$  and  $\Delta m_{ee}^2$. The covariance matrix, $V = V^{\rm{stat}} + V^{\rm{syst}}$, contains all sources of statistic and systematic errors affecting the experiment. The systematic error component was calculated from the full correlation matrix reported in \cite{Wong:2016}, including the signal, background and reactor core errors, and the total systematic uncertainties presented in figure 2 of Ref.~\cite{An:2015nua}. This assumption proved to be a reasonable approximation to the total systematic errors and correlations in the data set used for this analysis.

\begin{figure}[h!]
    \centering
    \includegraphics[scale=0.4]{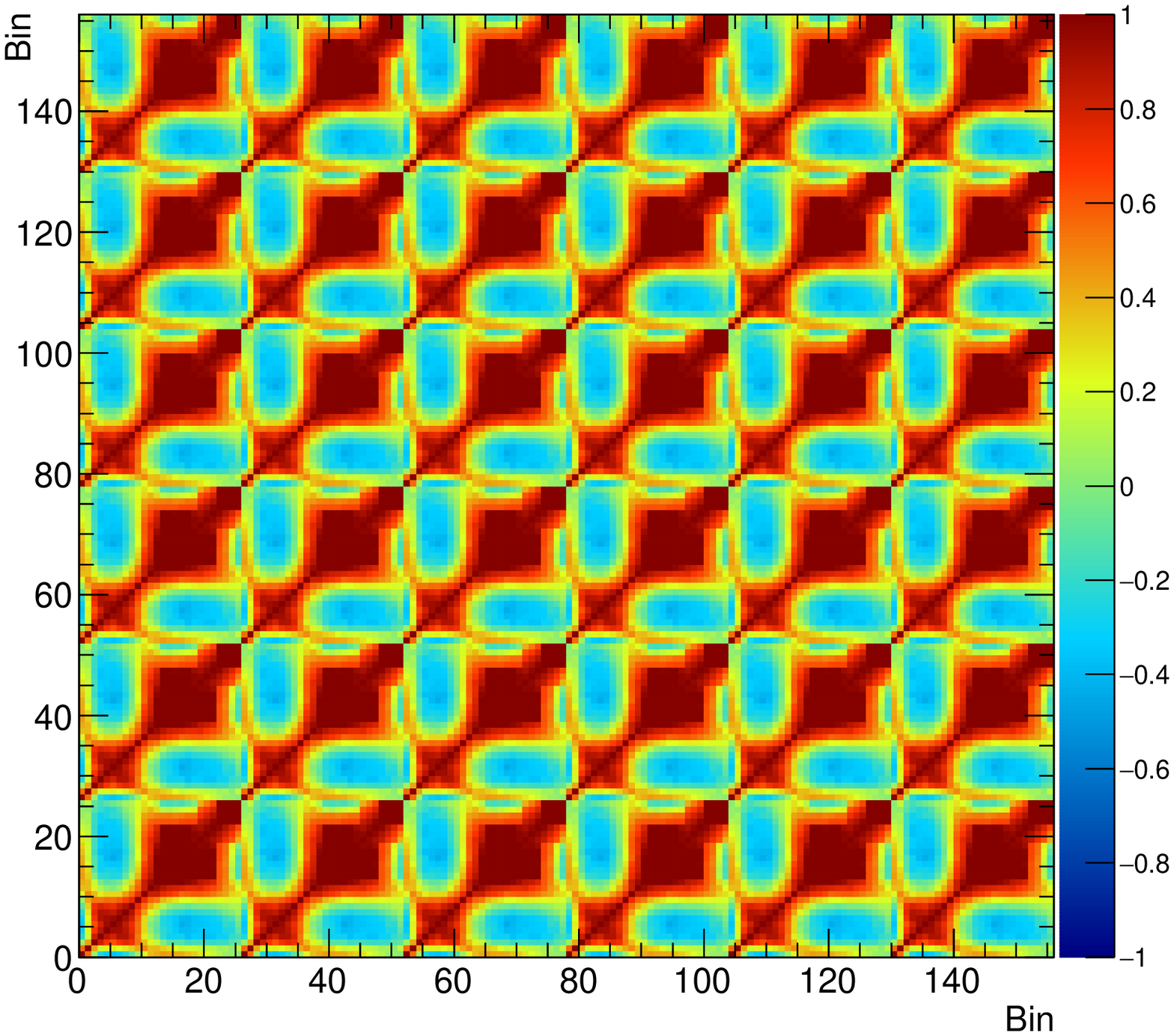}
    \includegraphics[scale=0.4]{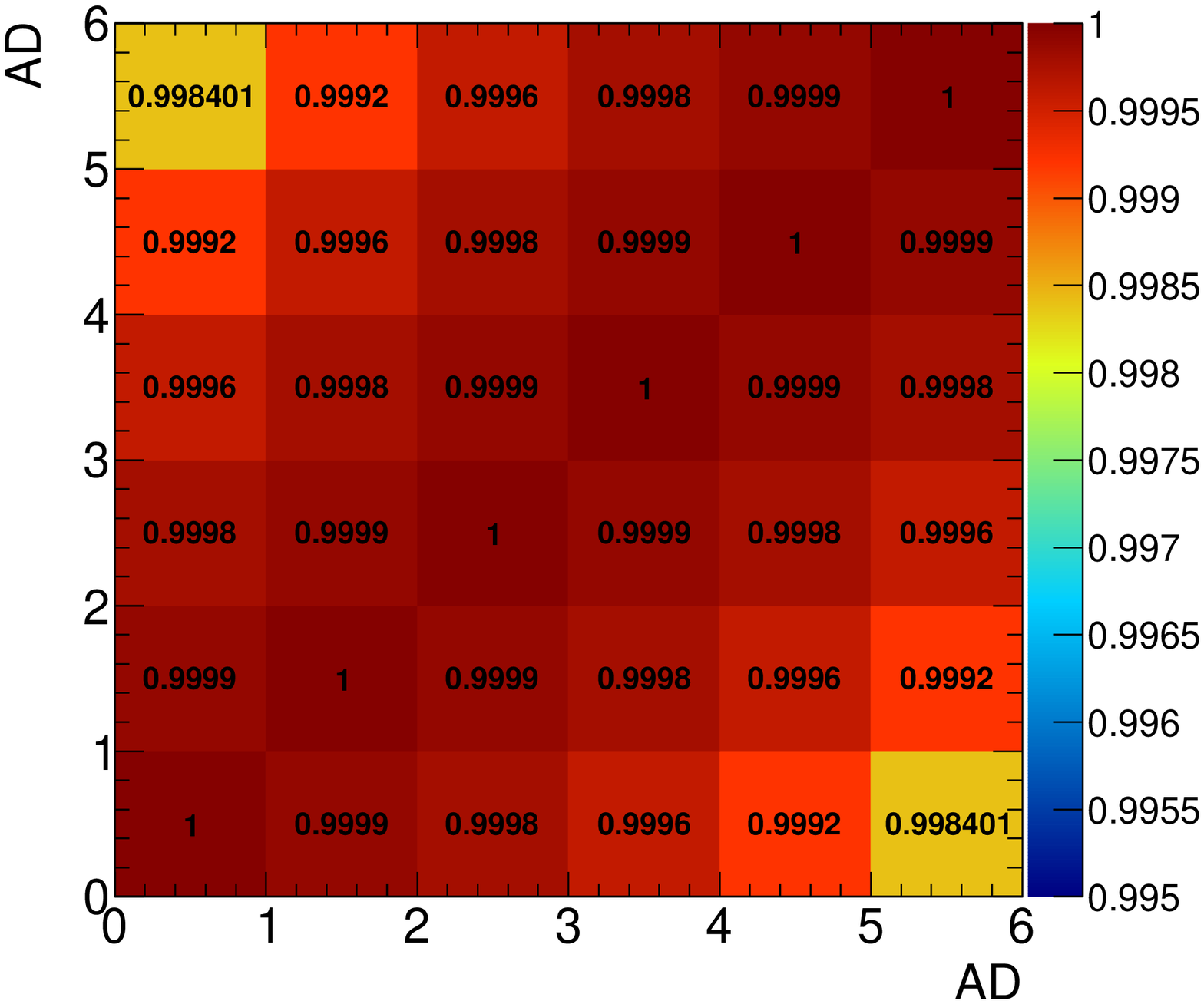}
    \includegraphics[scale=0.4]{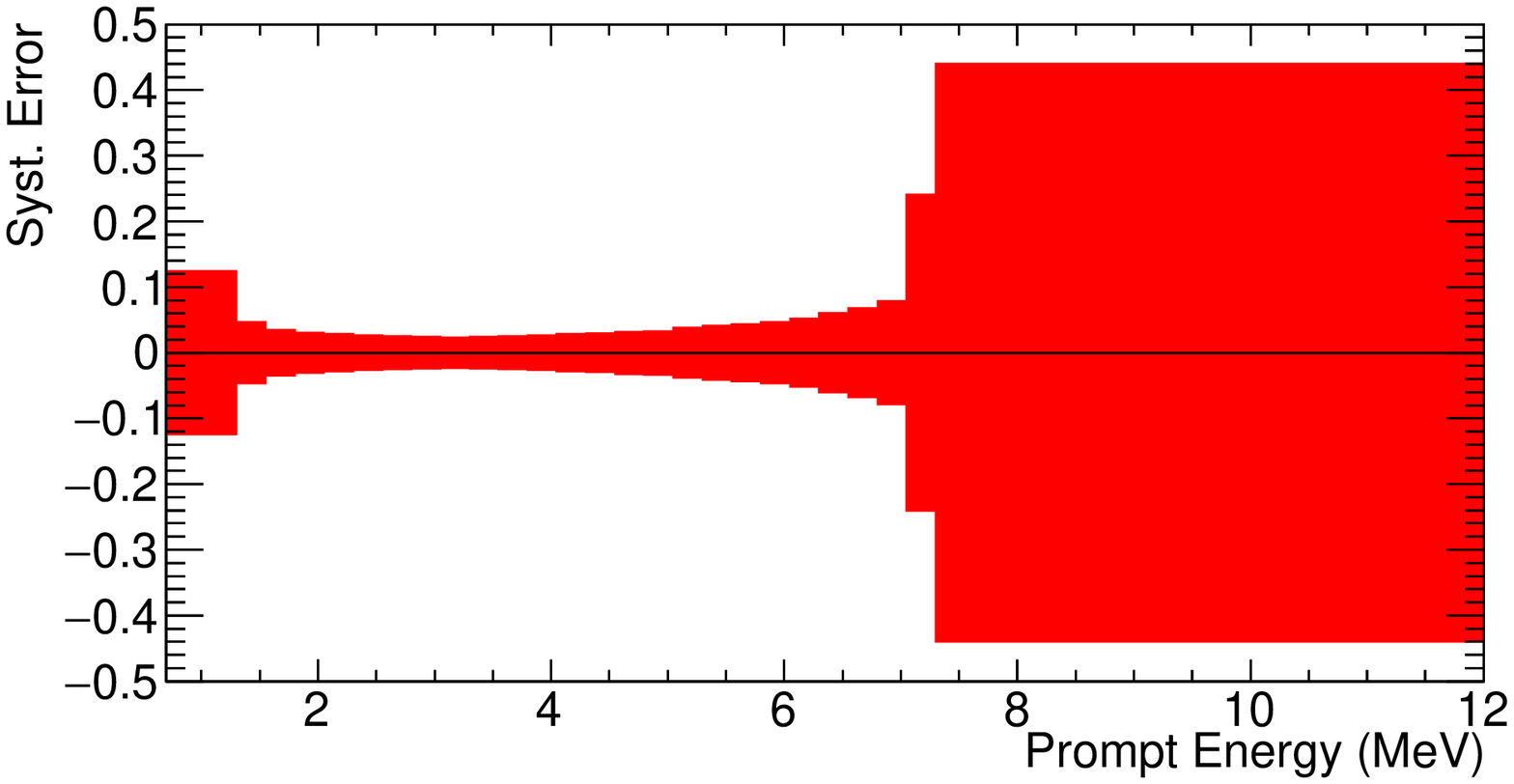}
\caption{\emph{Top:} Full systematic error (signal, background and reactor cores) correlation matrix used in the Daya Bay analysis with 26 bins per AD (signal, backgrounds and reactor cores). Each square block represents the correlation among the bins of two ADs. \emph{Center:} Correlation matrix for the 6 ADs. \emph{Bottom:} Total fractional systematic uncertainties in the prompt energy distribution, assumed equal for all ADs.\label{fig_DB_corrMatrix}} 
\end{figure}

The systematic error correlation matrix in \cite{Wong:2016} is a $222\times222$ matrix, since the energy spectrum in each AD used therein has 37 (non-uniform) bins in the energy range 0.7-12~MeV. However, the Daya Bay data set considered here used energy distributions with 26 bins, in the same energy range, hence having different bin boundaries (see Fig.~\ref{fig_rebin}). In order to cast this correlation matrix in the form a $156\times156$ matrix, we implemented a re-binning procedure based on the diagonal blocks (37$\times$37 bins per AD) of the original matrix in which we sampled energy distributions consistent with the original matrix \cite{devroye:1986}, and for each one, sampled $10^6$ energy values which were then filled into a re-binned histogram ($26\times26$ bins per AD). The resulting 1000 re-binned distributions were used to re-calculate the correlation matrix with the desired binning. The full 6 AD correlation matrix, $\rho^{\rm syst}$, was constructed assuming a correlation among the 6 detectors encoded in a $6\times 6$ matrix (see midle plot in Figure \ref{fig_DB_corrMatrix}) which was adjusted so that the overall features of the original matrix could be reproduced. Our re-binned correlation matrix is shown in the top panel of Figure \ref{fig_DB_corrMatrix}. 
The $156\times 156$ elements of the total systematic covariance matrix $V^{\rm syst}$ were then computed as
\begin{equation}\label{eq_systMatrix}
V^{\rm syst}_{ij} = \rho^{\rm syst}_{ij}
\times
\left(N^{exp}_i \;\sigma_{(i\;{\rm mod}\;26)}\right) \; \times \left(N^{exp}_j \sigma_{(j\;{\rm mod}\;26)}\right)
\end{equation}
(no summation over repeated indices) where the $\sigma_k$ , $k=1, \dots, 26$, are the total fractional systematic uncertainties, shown in the bottom panel of Figure \ref{fig_DB_corrMatrix}, extrapolated from \cite{An:2015nua} up to 12~MeV.
Despite being a rough approximation to the true error matrix used by the Collaboration, as we will see, our results agree reasonably well with those reported by Daya Bay.
%---------------------------------
%---------------------------------

\begin{figure}
\centering
{
\resizebox*{8.5cm}{!}{\includegraphics[angle=-90]{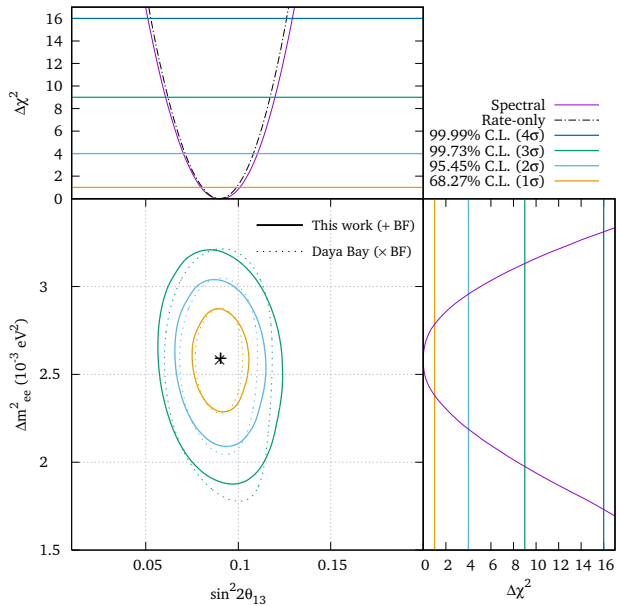}}}
\caption{Allowed regions in the ($\sin^2{2\theta}_{13},\Delta m_{ee}^2$) space at a 68.27\%, 95.45\%, 99.73\% C.L.~from our Daya Bay analysis. The best fit es marked by a `$+$' in the central plot, where the published Daya Bay contours \cite{An:2012eh} are also included. The marginalized $\Delta\chi^2$ over the oscillation parameters are also shown in the upper and right plots, including the rate-only result (dot-dashed line) for the mixing angle, and the 99.99\% C.L. line (solid dark blue).} \label{fig_DB_spec}
\end{figure}

We found that the oscillation parameters which best reproduce the data are $\sin^2{2\theta}_{13} =  0.091_{-0.009}^{+0.012}$ and $\Delta m_{ee}^2 = \left(2.60_{-0.22}^{+0.18}\right)\times 10{-3}$ eV$^2$ at 1$\sigma$ C.L., with $\chi_{\min}^2/\rm{NDF}=59.91/154$. This result is shown in Figure \ref{fig_DB_spec}, together with the 68.27\%, 95.45\% and 99.73\% C.L.~allowed regions for the oscillation parameters space. 
In the upper (right) panel of Figure~\ref{fig_DB_spec}, we show the $\Delta\chi^2$ marginalization over $\Delta m^2_{ee}$ ($\sin^2{2\theta}_{13}$), where the minimum of the curve ($\Delta\chi^2 = 0$) points to the best fit value. We have included here the result of the rate-only analysis (dash-dotted line in the top panel) for comparison purposes. In the marginalization plots, the horizontal (vertical) lines indicate the one-dimensional allowed regions for the two parameters at the same C.L.~as the 2D plot.

Although the best fit is very well recovered, our contours are slightly wider than the published ones towards (dotted line in Fig.~\ref{fig_DB_spec}) the higher $\sin^22\theta_{13}$ values, and shorter towards the lower $\Delta m^2_{ee}$. Despite our  efforts to reproduce the full covariance matrix for this measurement, several manipulations had to be implemented in order to re-bin the matrix and guarantee its positive definiteness,  which may have introduced distortions. Nonetheless, we consider that our result captures the main features of the analysis and gives a good approximation to the confidence regions for the parameters. A cross-check calculation using a $\chi^2$ with pull terms produced contours with similar characteristics. 
%---------------------------------
%---------------------------------
\subsubsection*{RENO}
Following \cite{Seo:2016uom} we define the $\chi^2$ statistic for the RENO spectral analysis as  
\begin{align}
    \chi^2 = &\sum_{i=1}^{N_{bins}} \frac{\left( O_i^{F/N} - T_i^{F/N} \right)^2}{U_i^{F/N}} \\
    &+ \sum_{d=N,F} \left(\frac{b^d}{\sigma_{bkd}^d}\right)^2 + \left(\frac{e}{\sigma_{scale}}\right)^2 + \left(\frac{\epsilon}{\sigma_{eff}}\right)^2.\nonumber
\end{align}
In this case, $O_i^{F/N}$ is the ratio of the observed IBD candidate events at the far detector over those observed at the near detector for the $i$-th energy bin; $T_i^{F/N} = T_i^{F/N}(b^d,e,\epsilon,\theta_{13},\Delta m_{ee}^2)$ is the corresponding ratio of expected events, and $U_i^{F/N}$ is the statistical uncertainty associated with $O_i^{F/N}$.

\begin{figure}
\centering
{
\resizebox*{8.5cm}{!}{\includegraphics[angle=-90]{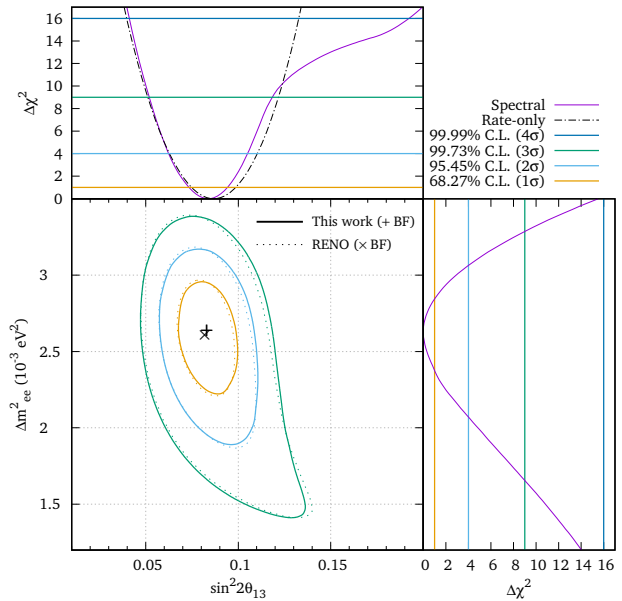}}}
\caption{Allowed regions in the ($\sin^2{2\theta}_{13},\Delta m_{ee}^2$) space at a 68.27\%, 95.45\%, 99.73\% C.L.~from our RENO analysis. The best fit es marked by a `$+$' in the central plot, where the published RENO contours \cite{Seo:2016uom} are also included. The marginalized $\Delta\chi^2$ over the oscillation parameters are also shown in the upper and right plots, including the rate-only result (dot-dashed line) for the mixing angle, and the 99.99\% C.L. line (solid dark blue).} \label{fig_RENO_spec}
\end{figure}

The best fit found after minimizing the $\chi^2$ statistic with respect to the oscillation parameters, marginalizing over  $b^d, e$ and $\epsilon$ (the pull term parameters), is $\sin^2{2\theta}_{13} = 0.083_{-0.012}^{+0.010}$ and $\Delta m_{ee}^2 = \left(2.64_{-0.27}^{+0.21}\right)\times10^{-3}$ eV$^2$ at 1$\sigma$ C.L., with $\chi_{\min}^2/\rm{NDF} = 20.92/25$. The results of this analysis are presented in Figure \ref{fig_RENO_spec}, where the allowed regions in the ($\sin^2{2\theta}_{13},\Delta m_{ee}^2$) parameter space are shown, together with the best fit point. As in the case for the Daya Bay analysis, the 1-dimensional marginalized distributions are also shown for each of the oscillation parameters, and the result obtained from the rate-only analysis (dashed line in top panel) for $\sin^2{2\theta}_{13}$ is included for comparison. 
We have also included the contours and best fit obtained by RENO  \cite{Seo:2016uom} in Fig.~\ref{fig_RENO_spec} (dotted line) which show good agreement with our analysis. 

The bottom-right plot in Figure \ref{fig_spectra} compares the FD data to the no oscillations and best fit predictions obtained from the measured spectrum at the ND. The agreement between the data and the best fit prediction is very good, and demonstrates that the near-to-far ratio technique used in the RENO analysis is insensitive to the presence of the 5 MeV bump.
%---------------------------------
%---------------------------------

\subsection{Combined Analysis}
Finally, we performed a combined analysis of the two data sets by considering the $\chi^2$ statistic
\begin{equation}\label{eq_COMB_chi2}
    \chi_{\rm{comb}}^2 = \Delta\chi_{\rm{Daya\,Bay}}^2 + \Delta\chi_{\rm{RENO}}^2.
\end{equation}
This definition will answer the specific question: how probable is it that both experimental results come from the same underlying oscillation model? \cite{Maltoni:2003cu}. Both, the rate-only and the spectral analysis were performed using the corresponding $\Delta\chi^2 = \chi^2\ - \chi_{\min}^2$ statistic appropriate for each case.

\begin{figure}
\centering
{
\resizebox*{8.4cm}{!}{\includegraphics[angle=-90]{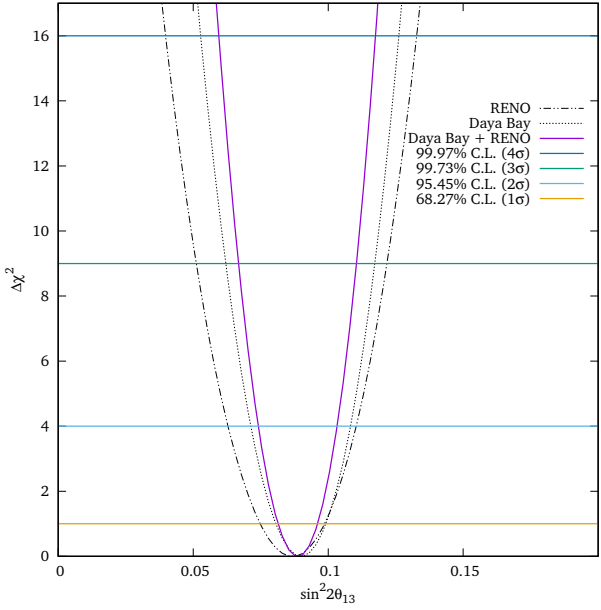}}}
\caption{Obtained $\Delta\chi^2$ distribution for $\sin^2{2\theta}_{13}$ from the Daya Bay + RENO rate--only combined analysis (full purple line), together with the results from Daya Bay (dotted line) and RENO (dash--dotted line), separately.} \label{fig_COMB_rate}
\end{figure}

For the combined rate-only analysis we found that the data is best described with a value of
$\sin^2{2\theta}_{13} = 0.088_{-0.006}^{+0.008}$.
This result is shown in Figure \ref{fig_COMB_rate}, where 
$\Delta\chi^2 = \chi_{\rm{comb}}^2-\chi_{\rm{comb}\,min}^2$ 
is plotted as a function of $\sin^22\theta_{13}$ (solid purple line). We also plot here the rate-only results from the independent analyses of Daya Bay (dotted line) and RENO (dash-dotted line). Horizontal colored lines are drawn to mark the allowed intervals for the parameter at 1-4$\sigma$ C.L., which are smaller for the combined analysis, indicating the expected enhancement in significance.
%---------------------------------
%---------------------------------

\begin{figure}
\centering
{
\resizebox*{8.5cm}{!}{\includegraphics[angle=-90]{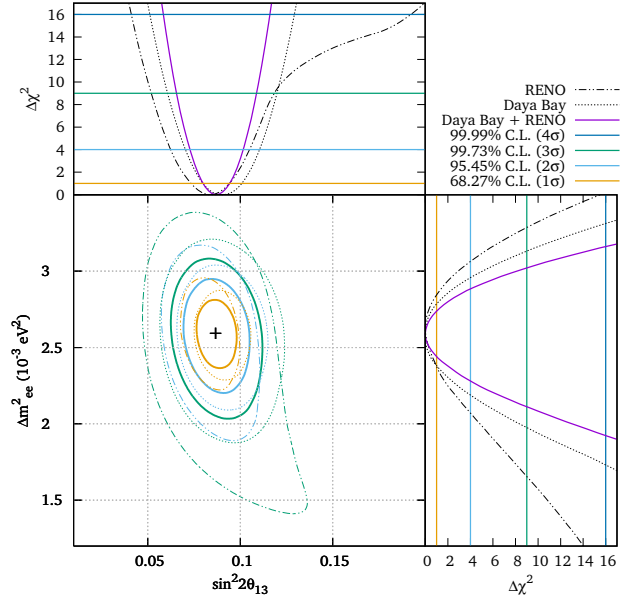}}}
\caption{2D regions ($1, 2$ and $3\sigma$) and 1D intervals ($1, 2, 3$ and $4\sigma$) allowed for the oscillation parameters obtained from the combination of Daya Bay and RENO data, using the spectral analysis. The best fit is marked by a `$+$' in the central plot. The results from the independent analyses (RENO dot-dashed; Daya Bay dotted) are shown for comparison.} \label{fig_COMB_spec}
\end{figure}

For the spectral combined analysis, the $\chi_{\rm{comb}}^2$ is also built as the sum of the corresponding statistics used for Daya Bay and RENO, as in (\ref{eq_COMB_chi2}). The minimization of such a function gives the values of the oscillation parameters which produce the best fit to the data: $\sin^2{2\theta}_{13} = 0.087^{+0.007}_{-0.008}$, 
$\Delta m_{ee}^2 = \left(2.59_{-0.14}^{+0.15}\right)\times10^{-3}$ eV$^2$  (1$\sigma$), with $\chi_{\rm{comb}\,min}^2/\rm{NDF} = 0.18/2$. 
Together with the best fit point, Figure \ref{fig_COMB_spec}  shows the 68.27\%, 95.45\% and 99.73\% C.L.~allowed regions in the studied parameter space. 

We also show in Figure \ref{fig_COMB_spec} (top and right panels) the 1-dimensional $\Delta\chi^2$ distributions for each oscillation parameter, marginalizing over the other one, where we have included the results obtained separately for Daya Bay (dotted line) and RENO (dash-dotted line). Clearly, the combined result (solid line) produces smaller intervals for $\sin^2{2\theta}_{13}$ and $\Delta m_{ee}^2$.

Finally, using the prescription described in Ref.~\cite{Maltoni:2003cu} we evaluate the compatibility between the two data sets, by calculating the parameter goodness (PG) as  the $\chi$-squared probability ${\rm Prob}(\chi_{\rm{comb}\,min}^2;P_c)$, where $P_c = 2$, is the number of parameters coupling the two data sets. In this case we obtain a compatibility of PG = 91.5\%. We note that the small discrepancies between our results for Daya Bay and the published ones, may be leading us to a different compatibility level from what could be obtained with the official analysis results. Our wider contour in $\sin^22\theta_{13}$ tends to increase the compatibility, while being smaller along $\Delta m^2_{ee}$ tends to reduce it. However, with regards to the question at the beginning of this section, the compatibility found here allows us to state that both, the Daya Bay and RENO data considered in this work, are well described by the same neutrino oscillation model, represented by (\ref{eq_survProb_reduced}), with the oscillation parameters found in the combined analysis.
%---------------------------------
%---------------------------------
%******************************************************************************
%******************************************************************************
\section*{Conclusions}
We have studied the neutrino oscillation analyses from two particular data taking periods of the Daya Bay and RENO experiments. We reproduced reasonably well the published rate-only and spectral analyses results from both experiments, obtaining, for the spectral analysis: 
\begin{align*}
\sin^2{2\theta}_{13} = 0.091_{-0.012}^{+0.009}, &\;\; 
\Delta m_{ee}^2 = \left(2.60_{-0.22}^{+0.18}\right)\times 10^{-3} \, \rm{eV}^2  \\ &{\rm [Daya \;Bay];}
\end{align*}
and
\begin{align*}
\sin^2{2\theta}_{13} = 0.083_{-0.012}^{+0.010}, &\;\;  
\Delta m_{ee}^2 = \left(2.64_{-0.27}^{+0.21}\right)\times10^{-3} \,  \rm{eV}^2   \\ &{\rm [RENO].}
\end{align*}
The spectral analysis of Daya Bay was the more challenging, considering our choice to use the full systematic error covariance matrix in 26 prompt positron energy bins in the definition of the $\chi^2$ statistic. This required the implementation of a statistical method to re-bin the correlation matrix found in a more recent publication by the collaboration. As a cross-check, we obtained very similar contours from a definition of the Daya Bay $\chi^2$ statistic using pull terms.
We were able to reproduce very closely all the results of the RENO spectral analysis, and verify that the Near/Far ratio technique makes the results insensitive to the presence of the 5~MeV bump.

A combined spectral analysis was carried out by defining a $\chi^2$ statistic as the sum of the Parameter Goodness (PG) $\Delta \chi^2$ for each data set, and extracting confidence regions around its minimum. We found that the values that best fit the data are 
\begin{align*}
\sin^2{2\theta}_{13} = 0.087&^{+0.007}_{-0.008}, \;\; 
\Delta m_{ee}^2 = \left(2.59_{-0.14}^{+0.15}\right)\times10^{-3} \,  \rm{eV}^2   \\ &{\rm [Daya \;Bay \;+ \;RENO]} 
\end{align*}
at 1$\sigma$ C.L.
The combined analysis provided more restricted allowed regions for the oscillation parameters, compared against the results from the two experiments separately, as expected, with an increase in the precision of the oscillation parameters from 30-40\%. Furthermore, we found the data sets considered here to be compatible at the 91.5\% level according to our analyses, despite small discrepancies with our result and the one published by Daya Bay.
%******************************************************************************
%******************************************************************************
\section*{Data Availability}
The data used to support the findings of this analyses are included within the article for better readability, and are also openly accessible in \cite{An:2013zwz,Seo:2016uom}.
%******************************************************************************
%******************************************************************************
\section*{Conflict of Interest}
The authors declare that they have no conflicts of interest.
%******************************************************************************
%******************************************************************************
\section*{Funding Statement}
This work was supported by Universidad del At\'antico through the grant ``Convocatoria Interna \emph{Impacto Caribe}'' No. CB71-CIC2014, and by Consejo Nacional de Ciencia y Tecnología (CONACyT, M\'exico) through SNI (Sistema Nacional de Investigadores).
%******************************************************************************
%******************************************************************************
\section*{Acknowledgements}
M.A.A.~and D.J.P.-T.~thank to the Organizing Committee of the \textit{VIII Encuentro Regional de Ciencias F\'isicas 2018}, Barranquilla, Colombia, in which preliminar results of this work where presented. M.A.A.~also thanks Instituto de Ciencias Nucleares for their hospitality during the partial realization of this work. A.A.A.-A.~acknowledges the hospitality of the Universidad del Atl\'antico during the time he spent working at the Physics Department.
%******************************************************************************
%******************************************************************************

\end{document}